\documentclass[balance=false,sigconf,authorversion,nonacm,anonymous=False]{acmart}

\usepackage{tikz}
\usetikzlibrary{positioning,arrows.meta,calc}
\makeatletter\g@addto@macro\UrlBreaks{\do\_\do\-}\makeatother
\DeclareUrlCommand\fld{\urlstyle{tt}}

\usepackage{caption}
\usepackage{multirow}
\usepackage{xcolor}
\usepackage{graphicx}

\usepackage{epstopdf}
\usepackage{url}
\usepackage{comment}
\usepackage{caption}
\usepackage{subcaption}
\usepackage{rotating}

\usepackage{tabularx}
\usepackage{booktabs}
\usepackage{multirow}
\usepackage{array}
\newcolumntype{P}[1]{>{\raggedright\arraybackslash}p{#1}}
\usepackage{makecell}

\usepackage{cuted}
\usepackage{capt-of}

\usepackage{enumitem}

\AtBeginDocument{%
  \providecommand\BibTeX{{%
  \normalfont 
  \kern-0.5em{\scshape i\kern-0.25em b}\kern-0.8em\TeX}
  }}
\setcopyright{acmlicensed}
\copyrightyear{2026}
\acmYear{2026}
\acmDOI{XXXXXXX.XXXXXXX}
\acmISBN{978-1-4503-XXXX-X/2018/06}

\graphicspath{{plots/}}
\begin{document}

\title{VeriPIE: Checkable Query Plans and Evidence-Guided Selection for Text-to-SQL}



\author{Piyush Jha}
\authornote{Both authors contributed equally to this research.}
\email{piyush@knowidea.dev}
\affiliation{%
  \institution{KNOWIDEA Technology}
  \city{Atlanta, GA}
  \country{USA}
}

\author{Cruise Song}
\authornotemark[1]
\email{csong@gatech.edu}
\affiliation{%
  \institution{Georgia Tech}
  \city{Atlanta, GA}
  \country{USA}
}

\author{Zhengyu Li}
\email{brian@knowidea.dev}
\affiliation{%
  \institution{KNOWIDEA Technology}
  \city{Atlanta, GA}
  \country{USA}
}

\author{Yatharth Sejpal}
\email{yatharth@knowidea.dev}
\affiliation{%
  \institution{KNOWIDEA Technology}
  \city{Toronto, ON}
  \country{Canada}
}

\begin{abstract}
Financial Artificial Intelligence (AI) governance and model risk management (MRM) emphasize independent validation, transparency, and effective challenge of AI-generated computations. Text-to-SQL translates natural-language questions into Structured Query Language (SQL), but queries that agree on one database can still express different computations. We present VeriPIE, a neurosymbolic framework combining language-model generation with symbolic checks. It ensembles direct SQL generation with QuerySketch, a structured plan language checked by a backend implemented in Lean, a theorem prover and programming language. Plan diagnostics guide repair before SQL generation, while execution agreement and bounded searches for distinguishing databases inform candidate selection. These checks expose malformed plans and hidden disagreement without proving agreement with the user's intent. In single-run experiments on the complete Spider development set, VeriPIE achieves 72.05\% test-suite accuracy (745/1,034), compared with DeepEye's 65.67\% (679/1,034). Its two-query shortlist covers 78.43\% (811 questions), compared with 74.66\% for DIN-SQL (772) and 69.05\% for DeepEye (714). On BIRD, VeriPIE reaches 68.90\% execution accuracy and 73.73\% two-query coverage. The central benefit is reviewable evidence about the generated computation: explicit plan checks, repair diagnostics, and concrete disagreements that support safer financial analysis.
\end{abstract}


\maketitle

\graphicspath{ {plots/} }

\section{Introduction}
\label{sec:intro}

Financial institutions operate under model risk management (MRM)
frameworks emphasizing validation, governance, and effective
challenge. For example, the Financial Stability Board
\footnote{\url{https://www.fsb.org/2024/11/the-financial-stability-implications-of-artificial-intelligence/}}
identifies AI-related model risk, data quality, and governance
as financial stability vulnerabilities, while the EU AI Act
\footnote{\url{https://eur-lex.europa.eu/eli/reg/2024/1689/oj}}
introduces risk-based requirements for covered AI systems.
These developments highlight the need for checkable components
in financial AI applications. This paper examines one such
component: text-to-SQL.

Text-to-SQL enables natural-language access to financial
databases through executable queries generated by large
language models (LLMs). Such queries may support portfolio
risk analysis, regulatory reporting, and transaction
reconciliation. However, an incorrect query can silently
omit exposures, double-count transactions, or miscalculate
financial aggregates while returning plausible results.
Traditional execution checks cannot reliably detect these
errors, and agreement among generated queries does not
establish computational correctness. This creates a distinct
validation challenge for financial AI: establishing what
a generated query computes and producing evidence that
independent validators can examine and challenge.

Execution agreement means that queries return the same results on the observed database. Consider ``Find accounts with no matching payment,'' with non-null account identifiers. A non-membership test (\texttt{NOT IN}) and a test for no matching row (\texttt{NOT EXISTS}) may agree when payment account identifiers contain no nulls, which represent missing or unknown values. If the payment account-identifier column permits nulls, adding one can cause \texttt{NOT IN} to exclude unmatched accounts because its condition becomes unknown, whereas \texttt{NOT EXISTS} still returns them. Agreement on current data therefore does not establish agreement on other valid databases.

VeriPIE addresses this gap by checking query plans, which describe database operations step by step, and making candidate agreement explicit. Building on DeepEye-SQL~\cite{deepeye}, it supports direct SQL generation and a plan language called QuerySketch. The latter uses Lean, a theorem prover and programming language, to check plans against the database schema: its tables, columns, and types. Failed checks return diagnostics that guide repair. Selection can combine execution agreement with structural checks and bounded searches, which look within a finite model for databases on which candidates disagree. These checks narrow uncertainty about the generated computation, but do not prove that it matches the user's intent.

The central idea is to connect candidate generation, plan checking, repair, and query selection through explicit evidence about each candidate. Figure~\ref{fig:pipeline} presents this workflow. Section~\ref{sec:method} describes how QuerySketch makes intermediate operations inspectable and how equivalence-aware grouping informs the final selection. Section~\ref{sec:results} evaluates the resulting framework on the complete Spider and BIRD development sets. Together, these contributions address the safety of generated financial queries by making their operations and the evidence behind selection open to review.

\section{Related Work}

Recent text-to-SQL systems divide inference into schema linking (matching question phrases to tables and columns), generation, revision, and selection. DeepEye-SQL follows this design, combining multiple generators with deterministic checks and candidate selection~\cite{deepeye}. VeriPIE extends generation with complementary strategies and a structured intermediate representation (IR) between natural language and SQL. It augments candidate selection with execution grouping, bounded verification, and ranking subject to additional checks. 

IRs reduce the gap between natural-language questions and executable SQL. SemQL and NatSQL simplify generation by abstracting selected SQL details~\cite{irnet,natsql}, whereas Query Plan Language (QPL) preserves query structure through explicit operator plans~\cite{qpl}. VeriPIE builds on QPL with QuerySketch, an extended plan language that is elaborated (resolved and type-checked against the schema) into a common representation of database operations in Lean~\cite{demoura2021lean4}. QuerySketch serves as a shared interface for validation, repair, SQL lowering (translation to executable SQL), and equivalence checking.

At selection time, execution voting chooses the most frequent result among sampled SQL candidates. This applies self-con\-sis\-tency~\cite{wang2023selfconsistency} to text-to-SQL~\cite{sqlpalm}. The method requires no reference query, but agreement on one database does not establish semantic equivalence, as distinct queries may return identical results on the observed instance yet diverge on another valid database. SpotIt mitigates this limitation through bounded verification with satisfiability modulo theories (SMT), a method for deciding logical constraints. It searches for counterexample databases---instances on which query results differ---within a fixed bound~\cite{barrett2018smt,spotit}. A discovered counterexample separates the queries. A completed check finding none supports equivalence only within its encoded model and bound. VeriPIE first ranks candidates by execution consensus, then can use plan checks, question-derived checks on expected structure and outputs (answer contracts), and bounded counterexamples to revise the ranking when evidence conflicts. If verification is unavailable or inconclusive, it falls back to the execution-based ranking.

LogicTrack similarly uses formal logic solvers to audit intermediate reasoning and guide backtracking, which revisits failed reasoning steps~\cite{logictrack2026}. VeriPIE applies a check-and-repair approach to executable query plans. TalkToModel offers a complementary natural-language interface for computing and discussing model explanations~\cite{talktomodel2023}. 

Our focus is query validity and candidate selection. Unlike approaches focused primarily on execution accuracy, VeriPIE emphasizes checkable computational evidence to support independent model validation in financial applications.

\section{The VeriPIE Approach}
\label{sec:method}
\begin{figure}[t]
\centering
\begin{tikzpicture}[font=\scriptsize,
 box/.style={draw=black!70,align=center,inner sep=1.5pt,minimum height=9mm},
 arr/.style={-{Stealth[length=1.1mm]},line width=.4pt}]
\node[box,text width=13.5mm] (gen) at (0,0) {Generate\\candidates};
\node[box,text width=15mm] (check) at (1.75,0) {Lean checks\\+ execution};
\node[box,text width=11mm] (repair) at (3.3,0) {Repair\\loop};
\node[box,text width=15.5mm] (bucket) at (4.95,0) {Equivalence\\buckets};
\node[box,text width=11mm] (select) at (6.65,0) {Final\\selection};
\draw[arr] (gen) -- (check);
\draw[arr] (check) -- (repair);
\draw[arr] (repair) -- (bucket);
\draw[arr] (bucket) -- (select);
\draw[arr] (repair.south) -- ++(0,-.3) -| (check.south);
\end{tikzpicture}
\caption{The VeriPIE workflow. Diagnostics guide repair, and revised candidates return to checking. Lean checks apply to QuerySketch plans. Execution agreement and bounded-equivalence evidence guide grouping and final selection.}
\Description{Generate candidates, apply Lean checks and execution, repair failures with a loop back to checking, group into equivalence-aware buckets, and select the final query.}
\label{fig:pipeline}
\end{figure}
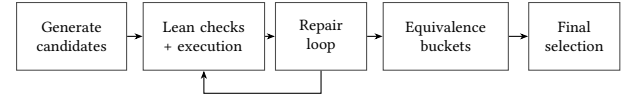

Given a question, database schema \(S\), and database instance \(D\), VeriPIE follows the five stages in Figure~\ref{fig:pipeline}: generate candidates, check and execute them, repair detected problems, group candidates, and select ranked answers. Direct SQL receives structural and execution checks. QuerySketch plans also pass through Lean-backed checking before SQL generation. Repair returns revised candidates to checking. Executable candidates are then organized into \emph{equivalence-aware buckets}, groups informed by execution agreement and bounded-equivalence evidence. Reference SQL is not used during inference. The main results report the combined system, which ensembles the Direct-SQL and Lean-Backed pipelines described below. Their executable candidates enter a common grouping and selection stage.

The workflow separates two questions: whether a candidate describes an executable computation, and whether that computation is a suitable answer. Plan checking and execution address the first question, while grouping and selection use the resulting evidence to address the second. This separation gives each stage a specific role. A type error can trigger repair before selection, while disagreement between executable candidates can inform ranking. The shared plan representation connects these stages without requiring the language model to produce a proof of its answer.

\subsection{Direct-SQL VeriPIE}

VeriPIE generates SQL through standard generation, question decomposition with self-correction, and schema-focused refinement, then applies verification-guided selection without an IR. Its bounded SQL verification uses SpotIt+, which is added by VeriPIE and is not part of native DeepEye. Each candidate undergoes deterministic structural and execution checks covering parsing, executability, schema references, output shape (the returned columns and their types), and question-relevant features such as aggregation, grouping, ordering, and limits. Candidates that pass these checks are partitioned into buckets according to their observed results on the target database. Buckets are ranked primarily by candidate support, the number of queries in each bucket, providing an initial execution-consensus ranking.

Answer contracts combine model proposals with rules grounded in the question. Each Lean generation proposes a partial contract. After deterministic parsing, a field is retained only when more than half of all generation samples agree on its value. The resulting contract is checked. Agreement does not prove correctness. Direct-SQL also uses handwritten safeguards. One flags replacing an explicitly supplied \texttt{SUM(x)}/\allowbreak\texttt{COUNT(*)} formula with \texttt{AVG(x)}: the former counts all rows in its denominator, whereas the latter excludes null values.

\begin{table*}[t]
\centering
\caption{Complete development-set results. Top-1 is first-query accuracy. Top-2 is coverage by either returned query (both percentages). Time is estimated mean seconds per question. Bold marks the highest accuracy or coverage.}
\label{tab:main}
\setlength{\tabcolsep}{5pt}
\renewcommand{\arraystretch}{1.12}
\begin{tabularx}{\textwidth}{@{}lXrrrrrr@{}}
\toprule
& & \multicolumn{3}{c}{Spider: test-suite evaluation} & \multicolumn{3}{c}{BIRD: execution evaluation} \\
\cmidrule(lr){3-5}\cmidrule(lr){6-8}
Method & Main mechanism & Top-1 & Top-2 & Time (s) & Top-1 & Top-2 & Time (s) \\
\midrule
LLM listwise selection & Joint ranking of the candidate list & 59.48 & 66.34 & 38.00 & 60.95 & 67.86 & 63.82 \\
MAC-SQL & Schema selection, decomposition, refinement & 64.99 & 68.09 & 5.26 & 63.89 & 67.01 & 19.06 \\
DeepEye & Execution agreement, confidence-aware judging & 65.67 & 69.05 & 35.44 & 68.45 & 73.66 & 60.58 \\
DIN-SQL & Decomposition and self-correction & 71.86 & 74.66 & 12.04 & 62.65 & 65.71 & 13.23 \\
\addlinespace[3pt]
VeriPIE & Plan checks, repair, equivalence-aware selection & \textbf{72.05} & \textbf{78.43} & 39.89 & \textbf{68.90} & \textbf{73.73} & 60.59 \\
\bottomrule
\end{tabularx}
\end{table*}

\subsection{Lean-Backed VeriPIE}

Lean-Backed VeriPIE modifies both the candidate-generation and selection stages of the DeepEye-SQL pipeline. Instead of generating database-engine-specific SQL, the language model produces multiple QuerySketch plans. QuerySketch is a QPL-derived IR whose parser, abstract syntax tree (AST), elaborator, and typed representation of database operations (the relational core) are implemented in Lean. The parser builds the AST, a tree describing the plan syntax. The elaborator resolves references and checks types. Its numbered steps scan tables, filter rows, project selected columns, join tables, aggregate values, sort or limit results, or apply set operations to earlier results. QuerySketch also extends QPL with typed arithmetic, conditional, string, and date expressions needed for BIRD and Spider.

For each question, the model receives the database schema, a compact QuerySketch syntax specification, and several examples. It returns one or more textual QuerySketch programs rather than Lean terms or SQL. Appendix~\ref{app:querysketch} illustrates how a two-step QuerySketch plan makes intermediate column dependencies explicit.

After generation, each textual QuerySketch candidate is passed with schema \(S\) to the Lean frontend. Let \(t_i\) denote the generated text, \(p_i\) its QuerySketch AST, and \(r_i\) a relational plan whose types reflect \(S\). The output \(s_i\) is SQL in dialect \(\delta\), the syntax used by the target database engine. The lowering pipeline is
\[
t_i \xrightarrow{\mathrm{parse}} p_i
    \xrightarrow{\mathrm{elaborate}_S} r_i
    \xrightarrow{\mathrm{lower}_\delta} s_i .
\]
Elaboration resolves schema references and checks types, column origins, aggregation, grouping, and output shape. Valid plans are lowered to SQL, while invalid plans produce structured diagnostics for model repair.

Every accepted QuerySketch plan has passed the frontend's checks for supported operations, schema references, and types. This gives a concrete validation guarantee at the plan level: acceptance depends on explicit checks, not the model's confidence in its own output. Rejected plans receive diagnostics that identify what must be corrected. This evidence remains useful even when two methods have similar benchmark accuracy. It does not establish that SQL lowering preserves every aspect of plan meaning or that the plan answers the question.

The typed relational core also supports bounded equivalence checking. Direct-SQL VeriPIE encodes pairs of SQL queries, whereas this variant encodes the relational plans elaborated from QuerySketch candidates. Sharing one representation between SQL lowering and equivalence encoding avoids reparsing dialect-specific SQL and reduces opportunities for semantic drift between execution and verification.

\subsection{Diagnostic Repair}
Appendix~\ref{sec:repair} gives a recorded repair example. Repair uses feedback from a failed check to revise the candidate that produced it. Feedback can concern plan structure and types, SQL execution, or the requested output. For example, an incompatible comparison calls for revisiting its operands, whereas a missing output column calls for revisiting the projection. These are different failures, even if both prevent a candidate from answering the question. The diagnostic makes the reason for revision explicit.

The return arrow in Figure~\ref{fig:pipeline} is essential: a revised candidate must pass through checking again. For QuerySketch, the revised text is parsed and elaborated before SQL generation. The generated SQL is then checked through execution. Repair thus connects candidate generation to validation rather than treating a model's revision as evidence of success. The repaired candidate contributes to selection through its checked plan and execution result, using the same grouping process as other executable candidates.

\subsection{Equivalence-Aware Grouping and Selection}
Candidates sampled from the same model can share errors. Moreover, distinct computations can agree on a particular database, so a popular execution result need not represent one consistent interpretation of the question. Bounded checking tests this apparent agreement rather than treating popularity as sufficient evidence.

For SQL candidates \(s_i\) and \(s_j\), execution buckets group those whose results agree on \(D\). These groups provide an initial execution-consensus ranking. Structural grouping matches normalized SQL or Lean plans while requiring compatible observed results. Bounded evidence can connect structurally different candidates or separate them with a counterexample. Write \(\llbracket s\rrbracket_D\) for the result of query \(s\) on database \(D\). Bounded verification searches for a database \(D'\) within a supported model and a finite search bound \(b\) such that
\[
\llbracket s_i\rrbracket_{D'}\ne\llbracket s_j\rrbracket_{D'}.
\]
A valid counterexample separates two candidates that may agree on \(D\), but it does not identify which candidate matches the user's intent. Grouping therefore informs selection alongside the question and structural checks, rather than replacing interpretation of the question.

A completed bounded check finding no counterexample provides evidence only within its encoded model and bound. The selector accepts the leading bucket when its execution support, structural checks, and verification evidence are consistent. Otherwise, it compares the strongest bucket representatives using question-specific checks, with execution voting as a fallback. Buckets are ranked first by sample count. For tied buckets, the earliest-generated valid candidate among the tied buckets is returned.

Lean checks plan structure and types. Bounded verification compares query results within its encoded semantics and bound. Unsupported constructs and timeouts remain unknown. The encoding must account for SQL behavior, including nulls, duplicate rows, ordering, and schema constraints.

For financial workflows, the advantage is a computation that can be examined before its output informs a decision. A reconciliation or transaction-review query can silently omit relevant records while returning plausible totals. QuerySketch exposes the operations, diagnostics identify rejected plans, and counterexamples reveal dependence on the observed data. Together, these provide concrete material for challenging a generated answer. Sampling more LLM candidates alone does not provide such checks. VeriPIE contributes a query-validation component to financial AI safety, with human interpretation still needed to assess the intended business meaning.

\section{Experimental Setup}
\label{sec:setup}

\subsection{Benchmarks and Baselines}
We use Spider 1.0~\cite{spider2018}, covering all 1,034 development questions across 20 databases, scored with the official test-suite evaluator~\cite{testsuite2020}. For BIRD~\cite{bird2023}, we evaluate all 1,534 development questions across 11 databases. We use the \href{https://huggingface.co/datasets/birdsql/bird_sql_dev_20251106}{November 2025 annotations} with the benchmark databases and official execution accuracy. Spider requires agreement with the reference across all test databases. BIRD checks the benchmark instance alone, so coincidental agreement can count as correct. The scores are not directly comparable across benchmarks.

All methods use \texttt{openai/gpt-5.6-luna} through OpenRouter. Table~\ref{tab:main} reports one evaluation run per system. Reasoning effort was medium and \texttt{max\_tokens} was 16,384. Temperature and \texttt{top\_p} used their default values. These are our local results, not published leaderboard scores. Under a 16-slot budget, combined VeriPIE uses eight Lean-backed candidates plus four standard, two decomposition/self-correction, and two schema-focused/refinement SQL candidates. Checking and repair work can differ. The Lean arm allowed one plan-repair round. Direct-SQL verification in the full-development replay used SpotIt+, with bound~2 and an 8-second timeout per checked pair.

We choose baselines that represent different ways to generate and select queries. DeepEye~\cite{deepeye} generates 16 direct-SQL candidates and selects by execution agreement and its native ranking. The Direct-SQL-only ablation reuses these candidates, changing selection alone. LLM-only listwise selection jointly ranks the saved DeepEye direct-SQL candidate pool, providing a comparison with model-based ranking over those candidates. It does not use the Lean candidate pool.

MAC-SQL (2025)~\cite{mac_sql} uses agents for schema selection, question decomposition, and execution-guided refinement, making it a comparison for modular generation and repair. DIN-SQL (2023)~\cite{din_sql} combines schema linking, query decomposition, SQL generation, and self-correction. Its use of NatSQL for complex queries also makes it relevant to intermediate-representation approaches. These methods provide established design comparisons. CHASE-SQL~\cite{chase_sql}, which combines diverse generation with learned pairwise selection, provides a newer direction for comparison.

\subsection{Evaluation Metrics}
Top-1 is the primary metric and scores the first returned query. Top-2 measures whether either returned query passes, rather than automatic identification of the correct answer. For DIN-SQL and MAC-SQL, the second query comes from an independent complete generation attempt (Appendix~\ref{app:baseline-protocol}). Reference SQL is used only for scoring. Mean latency is estimated from parallel execution logs per question when returning up to two candidates.

\section{Results and Analysis}
\label{sec:results}

\subsection{Main Results}
Spider provides the clearest improvement. VeriPIE answers 745 questions correctly at top-1, compared with DeepEye's 679, a net increase of 66 questions or 6.38 percentage points. Its top-2 shortlist covers 811 questions, compared with 714 for DeepEye and 772 for DIN-SQL, net increases of 97 and 39 questions. Relative to DIN-SQL, the top-1 difference is only two questions (0.19 points), while the top-2 difference is 3.77 points. The stronger shortlist result is therefore distinct from the small top-1 difference. On Spider, combining the paths improves top-1 accuracy by 3.68 percentage points over Direct-SQL alone and 0.87 over Lean-Backed alone (Table~\ref{tab:ablation}).

BIRD shows a smaller difference: VeriPIE exceeds DeepEye by 0.45 points in top-1 accuracy and 0.07 points in top-2 coverage. Within VeriPIE, the gap between top-1 and top-2 corresponds to 66 Spider questions and 74 BIRD questions for which the second candidate succeeds after the first fails. This leaves room to improve selection. In an interactive workflow, a knowledgeable user could inspect the alternative if the first answer appears wrong. This is a potential use of the shortlist rather than an evaluated user interface.

VeriPIE spends computation on plan checking, diagnostic repair, and bounded verification in addition to generation and execution. On Spider, estimated mean latency is 39.89 seconds, compared with 35.44 for DeepEye and 12.04 for DIN-SQL. DIN-SQL offers nearly the same top-1 accuracy with a shorter wait, while VeriPIE adds explicit checking evidence and broader shortlist coverage. On BIRD, VeriPIE and DeepEye have nearly identical latency (60.59 and 60.58 seconds). The design targets financial analysis and review, where an inspectable basis for accepting a query can matter alongside response time. These aggregate timings do not isolate the cost of individual checking stages.

\subsection{Interpreting the Benchmark Difference}
The scoring rules suggest one hypothesis for the different gains. If two candidates produce the same evaluated result on BIRD's database, choosing between them cannot change that execution score. Splitting their bucket can still change its support and allow a different bucket to win. Spider's test suite can distinguish candidates that agree on the original instance but differ on other databases. It can therefore reward a choice that BIRD's single-instance metric cannot distinguish. On the fixed combined pool, adding bounded evidence to structural selection improves Spider top-1 by 1.16 points (Appendix~\ref{app:ablation}), but does not establish this explanation for the benchmark difference. Reference-based bounded evaluation, as in SpotIt~\cite{spotit}, could test whether predictions differ in ways that execution accuracy misses.

In general, from an MRM perspective, identifying computational disagreements and producing reviewable counterexamples may be valuable even when benchmark accuracy improvements are modest.

\section{Conclusion}

We presented VeriPIE, a text-to-SQL pipeline that connects
generated answers to checkable evidence. Direct-SQL VeriPIE
combines execution agreement with structural checks and
bounded counterexamples. Lean-Backed VeriPIE adds QuerySketch
plans that are checked, repaired, and lowered through a typed
relational core. Their ensemble joins flexible SQL generation
with explicit plan validation. For financial AI safety and
model risk management (MRM), the contribution is an inspectable
computation that supports independent validation and effective
challenge: independent reviewers can examine which operations passed checks,
why a revision was needed, and where candidate answers disagree.
Such evidence can strengthen AI governance without establishing
semantic correctness or regulatory compliance.

Future work will evaluate financial reporting and regulatory
workloads and extend counterexample-guided judging, repair,
and efficient verification. Appendix~\ref{app:future}
details these directions.

\begin{acks}
We are grateful to Ava Pettit, Preetha Saha, Mingshu Li, and Dhagash Mehta of BlackRock, Inc. for their close collaboration on this work, including discussions that shaped its motivation in financial model risk management, guidance on the design of the evaluation, and detailed review of the manuscript.
\end{acks}
 
\bibliographystyle{ACM-Reference-Format}
\bibliography{sample-base}


\begin{thebibliography}{17}


\ifx \showCODEN    \undefined \def \showCODEN     #1{\unskip}     \fi
\ifx \showDOI      \undefined \def \showDOI       #1{#1}\fi
\ifx \showISBNx    \undefined \def \showISBNx     #1{\unskip}     \fi
\ifx \showISBNxiii \undefined \def \showISBNxiii  #1{\unskip}     \fi
\ifx \showISSN     \undefined \def \showISSN      #1{\unskip}     \fi
\ifx \showLCCN     \undefined \def \showLCCN      #1{\unskip}     \fi
\ifx \shownote     \undefined \def \shownote      #1{#1}          \fi
\ifx \showarticletitle \undefined \def \showarticletitle #1{#1}   \fi
\ifx \showURL      \undefined \def \showURL       {\relax}        \fi
\providecommand\bibfield[2]{#2}
\providecommand\bibinfo[2]{#2}
\providecommand\natexlab[1]{#1}
\providecommand\showeprint[2][]{arXiv:#2}

\bibitem[Barrett and Tinelli(2018)]%
        {barrett2018smt}
\bibfield{author}{\bibinfo{person}{Clark Barrett} {and} \bibinfo{person}{Cesare
  Tinelli}.} \bibinfo{year}{2018}\natexlab{}.
\newblock \showarticletitle{Satisfiability Modulo Theories}.
\newblock In \bibinfo{booktitle}{\emph{Handbook of Model Checking}},
  \bibfield{editor}{\bibinfo{person}{Edmund~M. Clarke},
  \bibinfo{person}{Thomas~A. Henzinger}, \bibinfo{person}{Helmut Veith}, {and}
  \bibinfo{person}{Roderick Bloem}} (Eds.). \bibinfo{publisher}{Springer
  International Publishing}, \bibinfo{pages}{305--343}.
\newblock
\urldef\tempurl%
\url{https://doi.org/10.1007/978-3-319-10575-8_11}
\showDOI{\tempurl}


\bibitem[de~Moura and Ullrich(2021)]%
        {demoura2021lean4}
\bibfield{author}{\bibinfo{person}{Leonardo de Moura} {and}
  \bibinfo{person}{Sebastian Ullrich}.} \bibinfo{year}{2021}\natexlab{}.
\newblock \showarticletitle{The {Lean 4} Theorem Prover and Programming
  Language}. In \bibinfo{booktitle}{\emph{Automated Deduction -- {CADE} 28}}
  \emph{(\bibinfo{series}{Lecture Notes in Computer Science},
  Vol.~\bibinfo{volume}{12699})}, \bibfield{editor}{\bibinfo{person}{Andr{\'e}
  Platzer} {and} \bibinfo{person}{Geoff Sutcliffe}} (Eds.).
  \bibinfo{publisher}{Springer}, \bibinfo{pages}{625--635}.
\newblock
\urldef\tempurl%
\url{https://doi.org/10.1007/978-3-030-79876-5_37}
\showDOI{\tempurl}


\bibitem[Eyal et~al\mbox{.}(2023)]%
        {qpl}
\bibfield{author}{\bibinfo{person}{Ben Eyal}, \bibinfo{person}{Moran Mahabi},
  \bibinfo{person}{Ophir Haroche}, \bibinfo{person}{Amir Bachar}, {and}
  \bibinfo{person}{Michael Elhadad}.} \bibinfo{year}{2023}\natexlab{}.
\newblock \showarticletitle{Semantic Decomposition of Question and {SQL} for
  Text-to-{SQL} Parsing}. In \bibinfo{booktitle}{\emph{Findings of the
  Association for Computational Linguistics: EMNLP 2023}}.
  \bibinfo{publisher}{Association for Computational Linguistics},
  \bibinfo{pages}{13629--13645}.
\newblock
\urldef\tempurl%
\url{https://doi.org/10.18653/v1/2023.findings-emnlp.910}
\showDOI{\tempurl}


\bibitem[Gan et~al\mbox{.}(2021)]%
        {natsql}
\bibfield{author}{\bibinfo{person}{Yujian Gan}, \bibinfo{person}{Xinyun Chen},
  \bibinfo{person}{Jinxia Xie}, \bibinfo{person}{Matthew Purver},
  \bibinfo{person}{John~R. Woodward}, \bibinfo{person}{John Drake}, {and}
  \bibinfo{person}{Qiaofu Zhang}.} \bibinfo{year}{2021}\natexlab{}.
\newblock \showarticletitle{Natural {SQL}: Making {SQL} Easier to Infer from
  Natural Language Specifications}. In \bibinfo{booktitle}{\emph{Findings of
  the Association for Computational Linguistics: EMNLP 2021}}.
  \bibinfo{publisher}{Association for Computational Linguistics},
  \bibinfo{pages}{2030--2042}.
\newblock
\urldef\tempurl%
\url{https://doi.org/10.18653/v1/2021.findings-emnlp.174}
\showDOI{\tempurl}


\bibitem[Guo et~al\mbox{.}(2019)]%
        {irnet}
\bibfield{author}{\bibinfo{person}{Jiaqi Guo}, \bibinfo{person}{Zecheng Zhan},
  \bibinfo{person}{Yan Gao}, \bibinfo{person}{Yan Xiao},
  \bibinfo{person}{Jian-Guang Lou}, \bibinfo{person}{Ting Liu}, {and}
  \bibinfo{person}{Dongmei Zhang}.} \bibinfo{year}{2019}\natexlab{}.
\newblock \showarticletitle{Towards Complex Text-to-{SQL} in Cross-Domain
  Database with Intermediate Representation}. In
  \bibinfo{booktitle}{\emph{Proceedings of the 57th Annual Meeting of the
  Association for Computational Linguistics}}. \bibinfo{publisher}{Association
  for Computational Linguistics}, \bibinfo{pages}{4524--4535}.
\newblock
\urldef\tempurl%
\url{https://doi.org/10.18653/v1/P19-1444}
\showDOI{\tempurl}


\bibitem[Hu et~al\mbox{.}(2026)]%
        {logictrack2026}
\bibfield{author}{\bibinfo{person}{Jingyu Hu}, \bibinfo{person}{Shu Yang},
  \bibinfo{person}{Weiru Liu}, {and} \bibinfo{person}{Di Wang}.}
  \bibinfo{year}{2026}\natexlab{}.
\newblock \showarticletitle{LogicTrack: Auditing Reasoning Trajectories of
  Large Language Models with Formal Logic Solvers}.
\newblock \bibinfo{journal}{\emph{arXiv preprint arXiv:2609.21492}}
  (\bibinfo{year}{2026}).
\newblock


\bibitem[Klopfenstein et~al\mbox{.}(2026)]%
        {spotit}
\bibfield{author}{\bibinfo{person}{Rocky Klopfenstein}, \bibinfo{person}{Yang
  He}, \bibinfo{person}{Andrew Tremante}, \bibinfo{person}{Yuepeng Wang},
  \bibinfo{person}{Nina Narodytska}, {and} \bibinfo{person}{Haoze Wu}.}
  \bibinfo{year}{2026}\natexlab{}.
\newblock \showarticletitle{{SpotIt}: Evaluating Text-to-{SQL} Evaluation with
  Formal Verification}. In \bibinfo{booktitle}{\emph{International Conference
  on Learning Representations}}.
\newblock
\urldef\tempurl%
\url{https://openreview.net/forum?id=iMkvR2ICSE}
\showURL{%
\tempurl}


\bibitem[Li et~al\mbox{.}(2026)]%
        {deepeye}
\bibfield{author}{\bibinfo{person}{Boyan Li}, \bibinfo{person}{Chong Chen},
  \bibinfo{person}{Zhujun Xue}, \bibinfo{person}{Yinan Mei}, {and}
  \bibinfo{person}{Yuyu Luo}.} \bibinfo{year}{2026}\natexlab{}.
\newblock \showarticletitle{{DeepEye-SQL}: A Software-Engineering-Inspired
  Text-to-{SQL} Framework}.
\newblock \bibinfo{journal}{\emph{Proceedings of the ACM on Management of
  Data}} \bibinfo{volume}{4}, \bibinfo{number}{3}, Article
  \bibinfo{articleno}{158} (\bibinfo{year}{2026}).
\newblock
\showeprint[arxiv]{2510.17586}~[cs.DB]


\bibitem[Li et~al\mbox{.}(2023)]%
        {bird2023}
\bibfield{author}{\bibinfo{person}{Jinyang Li}, \bibinfo{person}{Binyuan Hui},
  \bibinfo{person}{Ge Qu}, \bibinfo{person}{Jiaxi Yang},
  \bibinfo{person}{Binhua Li}, \bibinfo{person}{Bowen Li},
  \bibinfo{person}{Bailin Wang}, \bibinfo{person}{Bowen Qin},
  \bibinfo{person}{Ruiying Geng}, \bibinfo{person}{Nan Huo}, {et~al\mbox{.}}}
  \bibinfo{year}{2023}\natexlab{}.
\newblock \showarticletitle{Can {LLM} Already Serve as a Database Interface?
  {A} {BIg} Bench for Large-Scale Database Grounded Text-to-{SQLs}}.
\newblock \bibinfo{journal}{\emph{Advances in Neural Information Processing
  Systems}}  \bibinfo{volume}{36} (\bibinfo{year}{2023}),
  \bibinfo{pages}{42330--42357}.
\newblock


\bibitem[Pourreza et~al\mbox{.}(2024)]%
        {chase_sql}
\bibfield{author}{\bibinfo{person}{Mohammadreza Pourreza},
  \bibinfo{person}{Hailong Li}, \bibinfo{person}{Ruoxi Sun},
  \bibinfo{person}{Yeounoh Chung}, \bibinfo{person}{Shayan Talaei},
  \bibinfo{person}{Gaurav~Tarlok Kakkar}, \bibinfo{person}{Yu Gan},
  \bibinfo{person}{Amin Saberi}, \bibinfo{person}{Fatma Ozcan}, {and}
  \bibinfo{person}{Sercan~O Arik}.} \bibinfo{year}{2024}\natexlab{}.
\newblock \showarticletitle{{CHASE-SQL}: Multi-Path Reasoning and Preference
  Optimized Candidate Selection in Text-to-{SQL}}.
\newblock \bibinfo{journal}{\emph{arXiv preprint arXiv:2410.01943}}
  (\bibinfo{year}{2024}).
\newblock


\bibitem[Pourreza and Rafiei(2023)]%
        {din_sql}
\bibfield{author}{\bibinfo{person}{Mohammadreza Pourreza} {and}
  \bibinfo{person}{Davood Rafiei}.} \bibinfo{year}{2023}\natexlab{}.
\newblock \showarticletitle{{DIN-SQL}: Decomposed In-Context Learning of
  Text-to-{SQL} with Self-Correction}.
\newblock \bibinfo{journal}{\emph{Advances in Neural Information Processing
  Systems}}  \bibinfo{volume}{36} (\bibinfo{year}{2023}),
  \bibinfo{pages}{36339--36348}.
\newblock


\bibitem[Slack et~al\mbox{.}(2023)]%
        {talktomodel2023}
\bibfield{author}{\bibinfo{person}{Dylan Slack}, \bibinfo{person}{Satyapriya
  Krishna}, \bibinfo{person}{Himabindu Lakkaraju}, {and}
  \bibinfo{person}{Sameer Singh}.} \bibinfo{year}{2023}\natexlab{}.
\newblock \showarticletitle{Explaining machine learning models with interactive
  natural language conversations using TalkToModel}.
\newblock \bibinfo{journal}{\emph{Nature Machine Intelligence}}
  \bibinfo{volume}{5}, \bibinfo{number}{8} (\bibinfo{year}{2023}),
  \bibinfo{pages}{873--883}.
\newblock


\bibitem[Sun et~al\mbox{.}(2023)]%
        {sqlpalm}
\bibfield{author}{\bibinfo{person}{Ruoxi Sun}, \bibinfo{person}{Sercan~{\"O}.
  Arik}, \bibinfo{person}{Alex Muzio}, \bibinfo{person}{Lesly Miculicich},
  \bibinfo{person}{Satya Gundabathula}, \bibinfo{person}{Pengcheng Yin},
  \bibinfo{person}{Hanjun Dai}, \bibinfo{person}{Hootan Nakhost},
  \bibinfo{person}{Rajarishi Sinha}, \bibinfo{person}{Zifeng Wang}, {and}
  \bibinfo{person}{Tomas Pfister}.} \bibinfo{year}{2023}\natexlab{}.
\newblock \showarticletitle{{SQL-PaLM}: Improved Large Language Model
  Adaptation for Text-to-{SQL}}.
\newblock \bibinfo{journal}{\emph{arXiv preprint arXiv:2306.00739}}
  (\bibinfo{year}{2023}).
\newblock


\bibitem[Wang et~al\mbox{.}(2025)]%
        {mac_sql}
\bibfield{author}{\bibinfo{person}{Bing Wang}, \bibinfo{person}{Changyu Ren},
  \bibinfo{person}{Jian Yang}, \bibinfo{person}{Xinnian Liang},
  \bibinfo{person}{Jiaqi Bai}, \bibinfo{person}{Linzheng Chai},
  \bibinfo{person}{Zhao Yan}, \bibinfo{person}{Qian-Wen Zhang},
  \bibinfo{person}{Di Yin}, \bibinfo{person}{Xing Sun}, {et~al\mbox{.}}}
  \bibinfo{year}{2025}\natexlab{}.
\newblock \showarticletitle{{MAC-SQL}: A Multi-Agent Collaborative Framework
  for Text-to-{SQL}}. In \bibinfo{booktitle}{\emph{Proceedings of the 31st
  International Conference on Computational Linguistics}}.
  \bibinfo{pages}{540--557}.
\newblock


\bibitem[Wang et~al\mbox{.}(2023)]%
        {wang2023selfconsistency}
\bibfield{author}{\bibinfo{person}{Xuezhi Wang}, \bibinfo{person}{Jason Wei},
  \bibinfo{person}{Dale Schuurmans}, \bibinfo{person}{Quoc~V. Le},
  \bibinfo{person}{Ed~H. Chi}, \bibinfo{person}{Sharan Narang},
  \bibinfo{person}{Aakanksha Chowdhery}, {and} \bibinfo{person}{Denny Zhou}.}
  \bibinfo{year}{2023}\natexlab{}.
\newblock \showarticletitle{Self-Consistency Improves Chain of Thought
  Reasoning in Language Models}. In \bibinfo{booktitle}{\emph{The Eleventh
  International Conference on Learning Representations}}.
\newblock
\urldef\tempurl%
\url{https://openreview.net/forum?id=1PL1NIMMrw}
\showURL{%
\tempurl}


\bibitem[Yu et~al\mbox{.}(2018)]%
        {spider2018}
\bibfield{author}{\bibinfo{person}{Tao Yu}, \bibinfo{person}{Rui Zhang},
  \bibinfo{person}{Kai Yang}, \bibinfo{person}{Michihiro Yasunaga},
  \bibinfo{person}{Dongxu Wang}, \bibinfo{person}{Zifan Li},
  \bibinfo{person}{James Ma}, \bibinfo{person}{Irene Li},
  \bibinfo{person}{Qingning Yao}, \bibinfo{person}{Shanelle Roman},
  {et~al\mbox{.}}} \bibinfo{year}{2018}\natexlab{}.
\newblock \showarticletitle{{Spider}: A Large-Scale Human-Labeled Dataset for
  Complex and Cross-Domain Semantic Parsing and Text-to-{SQL} Task}. In
  \bibinfo{booktitle}{\emph{Proceedings of the 2018 Conference on Empirical
  Methods in Natural Language Processing}}. \bibinfo{pages}{3911--3921}.
\newblock


\bibitem[Zhong et~al\mbox{.}(2020)]%
        {testsuite2020}
\bibfield{author}{\bibinfo{person}{Ruiqi Zhong}, \bibinfo{person}{Tao Yu},
  {and} \bibinfo{person}{Dan Klein}.} \bibinfo{year}{2020}\natexlab{}.
\newblock \showarticletitle{Semantic Evaluation for Text-to-{SQL} with
  Distilled Test Suites}. In \bibinfo{booktitle}{\emph{Proceedings of the 2020
  Conference on Empirical Methods in Natural Language Processing (EMNLP)}}.
  \bibinfo{pages}{396--411}.
\newblock


\end{thebibliography}

\appendix
\section{Additional Examples and Future Work}
\subsection{QuerySketch and Repair Example}
\label{app:querysketch}
The \texttt{TopSort} operator sorts rows and keeps a requested number. For example, for non-null ages, ``return the name of one youngest singer'' becomes:

\noindent\begin{minipage}{\columnwidth}
\begin{verbatim}
#1 = Scan Table [ singer ]
     Output [ name, age ];
#2 = TopSort [ #1 ] Rows [ 1 ]
     OrderBy [ age ASC ]
     Output [ name ]
\end{verbatim}
\end{minipage}

The two steps make the requested computation explicit. The scan exposes both the name to return and the age needed for ordering. The second step selects a row using age and returns only its name. Column availability therefore matters between steps, even when a column is absent from the final answer. QuerySketch lets the checker examine these intermediate dependencies before generating SQL. It also gives repair a structured plan to revise when a reference or type is incompatible with the schema.

\label{sec:repair}
One recorded example asks, ``Show the stadium names without any concert.'' Initial QuerySketch plans failed type checking. Diagnostic repair produced a compatible \texttt{NOT EXISTS} plan that passed the official Spider test suite. The example links a detected plan error to a successful revision. 

\subsection{Agreement Beyond the Observed Database}
The account example in Section~\ref{sec:intro} illustrates why this extra evidence matters. Suppose the current payment table contains no null account identifiers. The two candidate queries can enter the same execution bucket because they return the same unmatched accounts. A valid database containing a null payment account identifier can distinguish the candidates even though their observed answers originally agreed. The counterexample supplies a concrete reason to reconsider the group: the apparent agreement depends on the current data. The question and schema then provide the context for choosing the intended treatment of unmatched accounts.

\subsection{Two-Answer Baseline Protocol}
\label{app:baseline-protocol}
DIN-SQL and MAC-SQL each obtain two answers through independent, complete runs of their own pipeline. Rank~1 retains the original final output. Rank~2 comes from another complete generation attempt, including the method's normal refinement steps. Top-2 scores success by either final output. A single reported evaluation thus includes two generation attempts per question for these baselines.

\subsection{Further Evaluation and Verification}
\label{app:future}
Future work will evaluate financial schemas and recurring reporting queries. We will also compare each system's BIRD predictions with reference SQL using bounded equivalence checking, reporting counterexamples, completed bounded checks, and unknown outcomes separately to study coincidental correctness.

For judging, an LLM or user could inspect competing outputs on a separating database and assess which matches the question. For repair, the generator could receive that database and the differing outputs to revise a candidate. These are proposed extensions rather than evaluated features.

Selective verification when leading buckets are close may reduce latency. Additional directions include comparisons with CHASE-SQL~\cite{chase_sql}, broader SQL dialect support, and proofs that plan transformations preserve query meaning.

\subsection{Generation and Selection Ablations}
\label{app:ablation}
An ablation removes or replaces a component to measure its contribution. We distinguish candidate generation from selection among candidates. Table~\ref{tab:ablation} reports Spider top-1 accuracy.

\begin{table}[htbp]
\centering
\caption{Spider top-1 accuracy. Higher values are better. All generation variants start with 16 slots. Selection variants reuse the combined 8+8 pool.}
\label{tab:ablation}
\setlength{\tabcolsep}{4pt}
\renewcommand{\arraystretch}{1.08}
\begin{tabularx}{\columnwidth}{@{}Xr@{}}
\toprule
Configuration & Top-1 (\%) \\
\midrule
\textbf{VeriPIE} & \textbf{72.05} \\
\addlinespace[3pt]
\multicolumn{2}{@{}l}{\emph{Candidate generation}} \\
Direct-SQL only (16 direct) & 68.37 \\
Lean-Backed only (16 Lean) & 71.18 \\
\addlinespace[3pt]
\multicolumn{2}{@{}l}{\emph{Selection on the same combined pool}} \\
Execution-only selection & 71.28 \\
Structural selection & 70.89 \\
\addlinespace[3pt]
\multicolumn{2}{@{}l}{\emph{Plan repair in the combined pipeline}} \\
Without Lean plan repair & 71.76 \\
\bottomrule
\end{tabularx}
\end{table}

\paragraph{Candidate generation.}
DeepEye generates 16 direct-SQL candidates and uses its native selector. The Direct-SQL-only ablation uses the same 16 candidates with structural and bounded-equivalence evidence, isolating a selection change without adding Lean-generated candidates. Lean-Backed VeriPIE instead generates 16 QuerySketch plans, checks them in Lean, permits repair, and lowers valid plans to SQL. Combined VeriPIE uses four standard direct-SQL candidates, two generated through decomposition and self-correction, two through schema-focused refinement, and eight Lean-backed candidates, followed by structural and bounded-equivalence selection. 

\paragraph{Selection and repair.}
The three selection configurations reuse the same combined 8+8 candidate pool and retain the shared question-based safeguards. Execution-only selection groups candidates by their observed database results and ranks by support. Structural selection groups matching normalized SQL or Lean plans, requiring compatible observed results. Structural plus bounded selection, used by combined VeriPIE, additionally uses verifier evidence to connect structurally different candidates and separate candidates when a counterexample is found. Execution-only selection is therefore not native DeepEye: its input includes Lean-generated candidates and it retains VeriPIE's question-based safeguards.

The repair ablation uses the original Lean plans before repair, retains invalid slots, and recomputes the relevant SQL checks. It changes candidate validity within the combined slot allocation rather than only replacing the selector on an unchanged executable pool.

\paragraph{Interpretation.}
Combining generation paths achieves the best Spider top-1 accuracy under the same 16-slot budget, exceeding both paths alone. On the fixed combined pool, structural plus bounded selection gains 1.16 percentage points over structural selection and 0.77 over execution-only selection. Structural selection alone does not improve on execution-only selection. Removing Lean plan repair costs 0.29 points. These comparisons support complementary generation and a benefit from bounded selection evidence in this experiment. 

\end{document}